# KICKER PULSERS FOR RECYCLER NOVA UPGRADES

Chris. C. Jensen, FNAL, Batavia IL, USA

*Abstract*

An upgrade of the Recycler injection kicker system required a faster rise time pulser. This system required a field rise and fall time of < 57 ns and a field flattop of 1.6 µs. This paper describes the variety of improvements made over the years that have resulted in this latest thyratron pulser. The effects of the trigger, the reservoir and the load impedance on delay and rise time will be discussed.

## OVERVIEW

These pulsed power supplies have a fairly high peak power (25 kV, 1 kA), modest repetition rate (15 Hz), very fast rise times (15 ns), very flat pulse tops (+/- 1% for 1.6 us), operate around the clock for many years with limited downtime (1 extended maintenance per year), and the magnets are in a significant radiation field (100 kRad / year). The rise time and flattop requirements are driven by the need for more beam power and maintaining low losses in the accelerator [1].

The topology chosen to meet all of these requirements is a floating switch in a remote service building and a resistive termination mounted at the magnet. The high radiation field and limited downtime require the pulser to be located outside the beam line enclosure. The tight flat top requirement requires a low loss pulse forming line. Additionally, the load must be located on the magnet to remove reflections between the load, cable and magnet that reduce flattop stability. To meet the high peak power and fast rise time requirements a thyratron switch was chosen. Low timing jitter and drift are required to maintain stable accelerator operation.

The magnet [2] and load [3] for this system have been previously described. The charging system for a different pulser was briefly described in [4] and [5] and further details are presented herein. The thyatron trigger system described previously [6] has been updated, but the filament and reservoir systems have never been described. Finally, measurements of current rise time for the complete system, without magnet, are presented.

## CABLE AND CHARGING SYSTEM

A pulse forming line (PFL) is used instead of a lumped pulse forming network to generate a very square pulse. Low dispersion is required to maintain the fast rise time for the cable run from the pulser to the magnet, about 45 m in this case, and to preserve the prompt fall time at the end of the pulse. The cable used has the same dimensions as RG220/U but with additional features. Two foil layers are bonded to the polyethylene core under the bare copper braid. This reduces dispersion at higher frequencies. The center conductor has stricter tolerance on dimensional stability and surface roughness to maintain tighter impedance control and increase voltage rating. High voltage pulse testing (60 kV) and impedance testing (±0.4 Ohm) are done on all cables to verify performance.

A two-step command resonant charging system is used. A capacitor, which stores slightly more energy than the PFL, is connected through a thyristor switch to a step up transformer, then through a diode to the PFL. The leakage inductance of the transformer and the capacitance determine the charging time. The capacitor charging supply fully charges the capacitor once a kicker fire is requested, typically to ~ 4 kV in ~ 30 ms. This reduces voltage stress on the capacitor and improves lifetime. The PFL is then charged, typically to ~ 55 kV in ~ 250 µs. This decreases the time there is voltage on the thyratron and means that the reservoir can be substantially increased while maintaining a low self triggering rate. That is a critical adjustment for achieving fast rise time as is shown later. The step up transformer is also used because of the limited voltage range of commercial charging supplies.

The resonant charging time is significantly shorter than in some other applications, but there have been no issues. The reduced time was due to accelerator transfer timing requirements and the jitter between different clock systems. A side benefit is the reduced size of the transformer. Balancing resistors of 6 Meg Ohm are used to insure equal voltage division across the thyratron gap during the rapid charging. The negative bias on the G2 voltage trigger must be low impedance to prevent thyratron self triggering.

## THYRATRON SUBSYSTEMS

### Filament and Reservoir

Stray capacitance from cathode to ground is one fundamental limit to the rise time. One contribution to this capacitance is the transformers used to couple filament, reservoir and trigger power to the cathode. At Fermilab, we have used high frequency (20 kHz) AC to couple power to a floating cathode since the early 1990s. This is done with two series transformers with a common, grounded intermediate winding. The intermediate winding is formed with a single turn of the improved RG220 cable going through the center of both transformers and the thyratron enclosure serving as the return. The ground conductors are removed from the cable, the center conductor is at ground potential, and the cathode enclosure forms an electric shield at high voltage. The capacitance of this arrangement is about 10 pF for each filament, reservoir and trigger transformer.

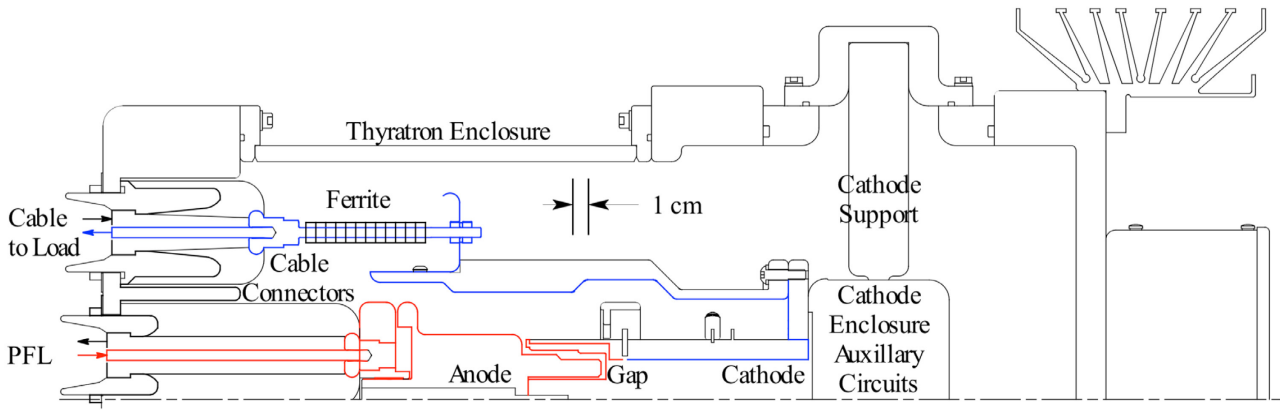

Figure 1. Sectional view of thyratron in enclosure. The thyratron side is axially symmetric with two inputs and outputs

The filament and reservoir power are created with a standard fixed frequency pulse width modulation control circuit applied to the primary of the first, low isolation voltage transformer. A regulated DC power supply is used in place of PWM regulation because there is no feedback from the high voltage side of the transformer. The duty cycle is approximately fixed but the control circuit keeps the transformers from saturating and provides over current protection. The resulting current waveform is roughly trapezoidal. The filament and reservoir voltage and current are measured at the low voltage side for monitoring.

Each thyratron's filament and reservoir current are measured at stamped nominal voltage and 60 Hz. The filament and reservoir voltage and current at 20 kHz are then measured at the cathode, without high voltage. The DC power supply voltage is adjusted until the 20 kHz rms current is the same as the 60 Hz rms current. This is the set point for the filament and the starting point for reservoir adjustment.

A recent improvement was to provide a slow feedback loop to regulate the filament and reservoir, as measured at the low voltage side. The average of the current squared is used to control the DC voltage of either the filament or the reservoir supply. This provides excellent stability of the reservoir temperature and is required to limit long-term drift in delay between trigger and thyratron turn on.

## Trigger

A pulsed, high current G1 trigger is now recommended by the manufacturer in place of a DC, low current bias for certain thyratrons. This pulsed current provides many more electrons to ionize the gas in the cathode area before the G2 trigger is applied [7], which should improve lifetime, jitter and rise time. A small DC current is applied continuously to G1 in case the G1 pulse is missing. The G1 and G2 pulses are both 2 μs long. A resistor outside the thyratron enclosure sets the G1 pulse current so it can be easily adjusted.

## Thyratron

The thyratron chosen for this application, E2V CX 2610, has two gaps with no drift space and has a maximum hold off voltage of 55 kV [8]. The cable is generally the primary limit for these types of systems, not the thyratron, because the cable has a reduced lifetime with pulsed operation > 60 kV.

## Enclosure

A geometry for the thyratron enclosure that has shown excellent performance is a re-entrant structure with the high voltage input and output on the same end. A tightly coupled path, from cable input through the thyratron and back to the cable output, reduces the inductance. The cathode auxiliary circuits have low capacitance to ground since they are not part of the low inductance main current path. This significantly decouples the series inductance and the parallel capacitance of the structure.

The enclosure is filled with fluid, both to cool the thyratron and provide high voltage insulation. Fluorinert™ FC-40 is used because it has a lower dielectric constant than mineral oil, $\varepsilon_r$ ~1.9 *vs*. $\varepsilon_r$ ~2.3, and lower boiling point, 165 C *vs*. 280 C. The fairly low boiling point means that generally no external pumping is required. The temperature and density differences in the fluid can generate natural convection cooling, transferring the heat from the thyratron to the thyratron enclosure. While Fluorinert™ is much more expensive than mineral oil it is non-flammable and there is no cost for pump or filter maintenance.

An outline drawing of the final enclosure for this pulser is shown in Figure 1. The cathode to ground capacitance is ~ 190 pF when filled with FC-40. The inductance was measured with the thyratron replaced with an aluminium rod the length of the thyratron but only 5 cm in diameter. This is roughly the size of the plasma inside the thyratron. The cable to anode inductance is ~ 45 nH, the cable to cathode inductance is ~ 50 nH and the dummy thyratron inductance is ~ 60 nH.

## MEASUREMENTS

*Delay Time*

The turn on delay of the thyratron is affected by gas pressure, G1 current trigger, G2 voltage trigger rise time and anode to cathode voltage. The G2 voltage trigger slew rate at the thyratron was fixed and was measured at > 100 kV/µs without high voltage. This is much higher than the manufacturer nominal, 10 kV/µs [8], and reduces jitter. The delay from G1 to G2 was changed from 400 ns to 1.5 µs to determine the effect on pulser delay with a G1 current of 16 A. The data sheet recommends a delay between 0.5 µs and 5 µs for this thyratron. The jitter from G2 trigger to conduction was less than 1 ns peak to peak over 5 minutes and about 0.2 ns rms, although the measurement resolution is only 0.1 ns. Table 1 shows several measurements of the delay.

The change in delay, or drift, with a G1 to G2 delay of 1000 ns, was measured several times over one week with the thyratron operating at a fixed repetition rate. The drift was 0.1 ns with the same 1 ns peak to peak jitter.

Operation with only a DC priming current on G1, of 110 mA, was measured but it is not recommended for long life. The delay increased by 38 ns and the jitter was about 3 ns peak to peak.

Table 1: Thyratron conduction delay as a function of G1 trigger to G2 trigger delay and reservoir, CX2610

| G1 to G2 Time Delay | Conduction Delay at Nom Res Power | Conduction Delay at 120 % Nom Res Power |
|---|---|---|
| 500 ns | 136.6 ns | 122.3 ns |
| 700 ns | 135.4 ns | 121.5 ns |
| 900 ns | 135.3 ns | 121.4 ns |
| 1100 ns | 136.2 ns | 122.0 ns |
| 1300 ns | 136.7 ns | 122.8 ns |

*Current Rise Time*

The turn on time of the thyratron is affected by gas pressure, G1 trigger current and anode to cathode voltage. The effect of these changes and the system impedance on the current rise time with a resistor load at the end of 19 m of cable, but without the magnet, was measured under several different operating conditions.

First, the reservoir was increased until the rise time did not substantially change. For this particular thyratron, that was about 120% of the reservoir power stamped on the tube. Increases in reservoir past 130% did significantly increase the self triggering rate.

Next, ferrite cores (13 mm OD, 8 mm ID, 6.5 mm long) were placed in series with each output connection. These shortened the rise time mainly by blocking the initial pulse from the first gap breaking down. Several different types, MnZn and NiZn, and core areas of ferrite were used. Changing the material type changed the shape of the rising edge of the current, however there was little change in the 10-90% rise time. Increasing the total area of cores had little effect after the initial improvement given by adding 12 cores.

Measurement of the current rise time for G1 to G2 trigger delays from 600 ns to 1200 ns indicated no change in rise time or wave shape. Changing from G1 DC priming of 110 mA to pulse of 9 A reduced current rise time by 2 ns. Increasing G1 from 9 A to 16 A made very little additional reduction.

The impedance of the system was changed between 50 Ohms, 25 Ohms and 16 2/3 Ohms. Simple simulations suggest that in all cases the system should have a rise time of less than 9 ns and that the 50 Ohm case should have a slower, over-damped wave shape. Unfortunately, these were not the results obtained during testing. The test data implies that the thyratron is the limiting factor in the rise time and that thyratron voltage collapse, not dI/dt, is the primary effect. Table 2 shows a summary of the rise time results.

Table 2: Current rise time as a function of operating conditions, CX2610 in low impedance housing

| Operating Condition | 5-95% Rise Time | 10-90% Rise Time |
|---|---|---|
| Nom Res Pwr, G1, 25 Ohm | 28.5 ns | 21 ns |
| 112% Res Pwr, G1, 25 Ohm | 24.5 ns | 19 ns |
| 120% Res Pwr, G1, 25 Ohm | 23 ns | 17 ns |
| 120% Res, G1, Ferrite, 25 Ohm | 16 ns | 12 ns |
| 120% Res, G1, Ferrite, 17 Ohm | 17.5 ns | 13 ns |
| 120% Res, G1, Ferrite, 50 Ohm | 17.5 ns | 13 ns |

## CONCLUSION

These systems have been working reliably for two years now, with over $10^9$ pulses accumulated. Very little tuning of timing has been needed to keep stable, low loss beam operation.

## ACKNOWLEDGEMENTS

This effort has been the result of years of work by many people. Special thanks Dirong Chen, Petar Dimovski, John Dinkel, George Krafczyk, Darren Qunell, Rob Reilly, David Tinsley and Steve Ward